\documentclass[aps,prl,preprint,groupedaddress,longbibliography,nofootinbib]{revtex4-1}

\usepackage{amsmath}
\usepackage{amssymb}
\usepackage{graphicx}

\begin{document}
\title{Reply to ``Comment on `Spin-dependent electron transmission model for chiral molecules in mesoscopic devices' ''}
\author{Xu~Yang}
\email[]{xu.yang@rug.nl}
\affiliation{Zernike Institute for Advanced Materials, University of Groningen, NL-9747AG Groningen, The Netherlands}

\author{Caspar~H.~van~der~Wal}
\affiliation{Zernike Institute for Advanced Materials, University of Groningen, NL-9747AG Groningen, The Netherlands}

\author{Bart~J.~van~Wees}
\affiliation{Zernike Institute for Advanced Materials, University of Groningen, NL-9747AG Groningen, The Netherlands}

\date{\today}

\maketitle 

The Comment by R.~Naaman and D.~H.~Waldeck addresses our recent publication \textit{``Spin-dependent electron transmission model for chiral molecules in mesoscopic devices~\cite{yang2019spin}''}. We believe that the Comment is largely based on misunderstandings of Ref.~\onlinecite{yang2019spin}, and it is important to clarify these in detail. Therefore, we provide here a point-by-point reply to the Comment and emphasize the important distinctions between the results obtained in the linear response regime and those obtained in the nonlinear regime, because both are experimentally observed using two-terminal electrical measurements.

\begin{itemize}
	\item[1.] Comment: \textit{``The paper published by Yang et. al.~\cite{yang2019spin} models spin transmission through chiral molecules in mesoscopic devices. Based on their model, they claim that spin selectivity in electron transport through chiral molecules, in the linear regime, cannot be measured by using a two-terminal device, unless a spin flip process occurs in the molecule.''} 
	
	Our remark: This summary of Ref.~\onlinecite{yang2019spin} is incorrect. First, the spin-flip reflection is directly related to the presence of spin-polarized transmission. Second, the spin-polarized transmission cannot be detected using a two-terminal electrical measurement in the linear response regime, regardless of the presence of spin-flip reflection.
	
	\item[2.] Comment: \textit{``Their simplified, two-terminal model assumes that charge is injected from a source electrode, transits through a chiral molecule and a ferromagnet, and is collected at a drain electrode. In this treatment, the ferromagnet transmits a given spin and reflects the other; but there is no dissipation in the ferromagnet. While the conclusions drawn by the authors may be consistent with the simplified model, the model itself is not realistic enough to account for experiments.''}
	
	Our remark: We indeed consider a two-terminal model but it does not involve simplifications for the linear response regime. Note that the role of the source and drain electrodes are interchangeable in the linear response regime because of microscopic reversibility. We have included in the model electron reflections at all interfaces, including the ferromagnet and the electrodes. The ferromagnet is characterized by a spin polarization parameter, which can be tuned from $0$ to $1$. The conclusions of the model are valid for all polarization values. 
	
	\item[3.] Comment: \textit{``Theoretical models for the CISS effect, in two contact spin measurements, exist in the literature already, and the conditions for observing spin polarization have been discussed in detail. As an example, consider the work by Matityahu et. al. which states: ``When the helix is connected to two one-dimensional single-mode leads, time-reversal symmetry prevents spin polarization of the outgoing electrons. One possible way to retrieve such a polarization is to allow leakage of electrons from the helix to the environment, via additional outgoing leads.'' ''}
	
	Our remark: How a spin polarization can be generated by a chiral molecule has indeed been discussed in many publications, including Refs.~\onlinecite{matityahu2016spin,medina2015continuum,nurenberg2019evaluation} mentioned in the Comment. This is also pointed out in Ref.~\onlinecite{yang2019spin}. However, Ref.~\onlinecite{yang2019spin} addresses a completely different issue, which is how such a spin polarization can be detected as a charge signal in transport experiments in the linear response regime. To our best knowledge, this issue is only addressed in one other publication~\cite{dalum2019theory}, which appeared after Ref.~\onlinecite{yang2019spin}.
	
	\item[4.] Comment: \textit{``In other words, dephasing acts to create asymmetry in the transmission amplitude for spin up versus spin down, and it breaks Onsager's reciprocity relation.''}
	
	Our remark: While dephasing indeed creates transmission asymmetry for opposite spins, it does not break reciprocity. The Onsager's relation is a thermodynamical theorem, and it holds in the presence of dephasing, see for example Ref.~\onlinecite{buttiker1986role}.
	
	\item[5.] Comment: \textit{``For example, Buttiker~\cite{sanchez2004magnetic} showed how asymmetry arises for magnetoconductance in a two terminal device. The combination of interactions with a bath and	the large electric fields at interfaces (typical of CISS experiments) can result in the observed asymmetry.''}
	
	Our remark: This is correct, but the asymmetry can only occur outside the linear response regime, i.e. away from zero bias by at least $V=k_B T/e$~\cite{datta1997electronic}.
	
	\item[6.] Comment: \textit{``In addition, we note that spin-selective backscattering, as an explanation for the spin selectivity, was also discussed previously~\cite{medina2015continuum} and even used to analyze for the extent of spin flipping in experiments~\cite{nurenberg2019evaluation}.''} 
	
	Our remark: This is the same issue as explained in the above Point 3. Refs.~\onlinecite{medina2015continuum,nurenberg2019evaluation} discuss how a spin-polarized current can be generated by chiral molecules, but not how it can be electrically detected in a charge transport experiment.
	
	\item[7.] Comment: \textit{``To summarize, two-terminal models have been discussed before, and it was shown that CISS can be observed if dissipation or a combination of non-linearity and dissipation are included.''}
	
	Our remark: This is incorrect, see the previous discussions (Points 3-6). 
	
	\item[8.] Comment: \textit{``The origin of the nonlinearity, to which we refer, is important to clarify. The simplified model used by Yang et al, presents the linear approximation for the conduction, but it does not relate to the actual parameters characterizing the CISS measurements and could prove misleading to some readers.''} 
	
	Our remark: This is incorrect. The model in Ref.~\onlinecite{yang2019spin} is not simplified in the linear response regime, and the results are strict. The model does not intend to present descriptions for the conduction beyond this regime. The model is very relevant for actual CISS measurements since for several of those, such as the ones shown in Fig.~1 in the Comment, the linear response regime can be clearly identified.
	
	\item[9.] Comment: \textit{``For charge moves through a system which is smaller, in dimension, than the screening length, the transport does not depend linearly on the field applied~\cite{ben1994nonlinear,middleton1993collective}. Because the chiral molecules studied, in all the works cited in Ref.~\onlinecite{yang2019spin}, are on the scale of few nanometers, upon applying an electric potential the typical field is of the order of $10^8$ V/m. Consequently, the electronic states in the molecules ‘mix’; and the electric field has two contributions: mixing of zeroth-order states by the Stark effect and driving current via the potential drop, conduction. For an example of a model based treatment, see the recent work by Michaeli~\cite{michaeli2019voltage}. This limit is different from most conduction studies of mesoscopic structures.''} 
	
	Our remark: This is irrelevant to the linear response regime. Electron transport is driven by a difference, or a gradient, of the electrochemical potential. In our case it is the difference in Fermi levels in the two electrodes. What the corresponding electric field (distribution) is depends on the electrostatic screening properties of the device, and indeed on whether the device length is shorter or longer than the screening length. In the linear response regime, the calculations of electron transport do not require the (self-consistent) calculations of electrostatic potentials and fields. Outside the linear response regime, though, the effects described in the Comment can indeed play an important role.
	
	\item[10.] Comment: \textit{``The non-linearity of the conduction is readily apparent in experiments with two contacts that have already been published. For example, Figure~1A presents the current versus potential curves, that were measured in a magnetic conducting probe AFM configuration and Figure 1B shows the corresponding plot of the conductance versus the applied potential. Figure 1C shows the spin polarization as a function of the applied potential, which is extracted from the measurements shown in Fig.~1A. Note that these data are obtained from “two contact experiments” that have been presented in figures 2 and 3 of a paper~\cite{kiran2016helicenes}, referred to as reference 6 by Yang et al.~\cite{yang2019spin}. The nonlinear response is apparent both in the current dependence on the voltage (Fig 1A), as well as in the other curves.''}
	
	Our remark: Fig.~1A and Fig.~1B (see in the Comment) greatly help us to clarify our point: In all the nonlinear curves in these figures, the linear response regime can be clearly identified, and here it is roughly within $\pm 0.05$~V. According to Ref.~\onlinecite{yang2019spin}, the red and blue curves in both Fig.~1A and Fig.~1B should overlap within this bias range, but they do not. More generally, for any CISS measurement using the magnetic conducting AFM technique, the two (averaged) $I$-$V$ curves should have the same slope at zero bias, and the two (averaged) $dI/dV$ curves should have the same value at zero bias. We emphasize that these requirements originate from the fundamental microscopic reversibility and the laws of thermodynamics. The departures from these requirements as shown in Fig.~1A and Fig.~1B, we think, may be related to the statistical approaches used in these experiments.
	
	\item[11.] Comment: \textit{``To illustrate the nonlinearity more clearly, Figure~1D shows a plot of the data from Fig.~1A on a semi-log graph. His plot reveals the exponential growth of the current at low voltage and the deviation of the currents from each other at higher voltages.''}
	
	Our remark: Figure~1A and Fig.~1D do not represent the same data, because (1) At a bias of $1$~V, the two curves in Fig.~1A reach values of about $0.8$~nA and $0.3$~nA, respectively, whereas the two curves in Fig.~1D reach about $1.0$~nA and $0.4$~nA; (2) At zero bias, the two curves in Fig.~1A have different slopes, which is also shown by a difference of $dI/dV$ values of a factor of two to three in Fig.~1B. This must result in a vertical shift between the curves in Fig.~1D over the entire range, but it is not present at low bias. Also, the labeling of the two curves, in the sense that whether it is the $H_{\text{DOWN}}$ or the $H_{\text{UP}}$ curve that gives higher current, is not consistent in Fig.~1A and Fig.~1D. We therefore will not comment on Fig.~1D.  
	
	\item[12.] Comment: \textit{``The spin polarization changes dramatically at low potentials; it is basically zero at very low fields and increases as the electric field approaches a maximum of~$\sim5\times10^8$~V/m. This observation, which is apparent in most current vs. voltage curves cited in ref.~\onlinecite{yang2019spin}, shows that the simplified model developed in ref.~\onlinecite{yang2019spin} is not relevant to the measurements.''}
	
	Our remark: This description is not inconsistent with our model. According to Ref.~\onlinecite{yang2019spin}, the polarization calculated as in Fig.~1C should be zero in the linear response regime, and then it may increase with increasing bias. Both curves in Fig.~1C indeed show zero spin polarization at zero bias, which proves the relevance of our model to actual measurements. However, it is unclear to us how the data points at zero bias in Fig.~1C have been obtained.
	
	\item[13.] Comment: \textit{``In summary, the model presented in ref.~\onlinecite{yang2019spin} oversimplifies; it fails to include the dissipation processes occurring at room temperature and it considers a linear limit that is not valid for the measurements on the CISS effect.''}
	
	Our remark: This conclusion is incorrect. Ref.~\onlinecite{yang2019spin} considers only the linear response regime which is clearly observed in experiments such as the one shown in Fig.~1 in the Comment, and therefore it is very relevant to actual measurements. Within the linear response regime the model is not simplified. 

\end{itemize}

We emphasize again that Ref.~\onlinecite{yang2019spin} is intended to raise awareness of the consequences of fundamental symmetries and limitations of certain electrical measurement geometries. It also highlights the differences between linear and nonlinear regimes. An extension of our model shows that a magnetoresistance in the two-terminal geometries discussed here can indeed be observed in the nonlinear regime~\cite{yang2019detecting}.


\begin{thebibliography}{13}%
	\makeatletter
	\providecommand \@ifxundefined [1]{%
		\@ifx{#1\undefined}
	}%
	\providecommand \@ifnum [1]{%
		\ifnum #1\expandafter \@firstoftwo
		\else \expandafter \@secondoftwo
		\fi
	}%
	\providecommand \@ifx [1]{%
		\ifx #1\expandafter \@firstoftwo
		\else \expandafter \@secondoftwo
		\fi
	}%
	\providecommand \natexlab [1]{#1}%
	\providecommand \enquote  [1]{``#1''}%
	\providecommand \bibnamefont  [1]{#1}%
	\providecommand \bibfnamefont [1]{#1}%
	\providecommand \citenamefont [1]{#1}%
	\providecommand \href@noop [0]{\@secondoftwo}%
	\providecommand \href [0]{\begingroup \@sanitize@url \@href}%
	\providecommand \@href[1]{\@@startlink{#1}\@@href}%
	\providecommand \@@href[1]{\endgroup#1\@@endlink}%
	\providecommand \@sanitize@url [0]{\catcode `\\12\catcode `\$12\catcode
		`\&12\catcode `\#12\catcode `\^12\catcode `\_12\catcode `\%12\relax}%
	\providecommand \@@startlink[1]{}%
	\providecommand \@@endlink[0]{}%
	\providecommand \url  [0]{\begingroup\@sanitize@url \@url }%
	\providecommand \@url [1]{\endgroup\@href {#1}{\urlprefix }}%
	\providecommand \urlprefix  [0]{URL }%
	\providecommand \Eprint [0]{\href }%
	\providecommand \doibase [0]{http://dx.doi.org/}%
	\providecommand \selectlanguage [0]{\@gobble}%
	\providecommand \bibinfo  [0]{\@secondoftwo}%
	\providecommand \bibfield  [0]{\@secondoftwo}%
	\providecommand \translation [1]{[#1]}%
	\providecommand \BibitemOpen [0]{}%
	\providecommand \bibitemStop [0]{}%
	\providecommand \bibitemNoStop [0]{.\EOS\space}%
	\providecommand \EOS [0]{\spacefactor3000\relax}%
	\providecommand \BibitemShut  [1]{\csname bibitem#1\endcsname}%
	\let\auto@bib@innerbib\@empty
	\bibitem [{\citenamefont {Yang}\ \emph {et~al.}(2019)\citenamefont {Yang},
		\citenamefont {van~der Wal},\ and\ \citenamefont {van Wees}}]{yang2019spin}%
	\BibitemOpen
	\bibfield  {author} {\bibinfo {author} {\bibfnamefont {Xu}~\bibnamefont
			{Yang}}, \bibinfo {author} {\bibfnamefont {Caspar~H}\ \bibnamefont {van~der
				Wal}}, \ and\ \bibinfo {author} {\bibfnamefont {Bart~J}\ \bibnamefont {van
				Wees}},\ }\bibfield  {title} {\enquote {\bibinfo {title} {Spin-dependent
				electron transmission model for chiral molecules in mesoscopic devices},}\
	}\href@noop {} {\bibfield  {journal} {\bibinfo  {journal} {Physical Review
			B}\ }\textbf {\bibinfo {volume} {99}},\ \bibinfo {pages} {024418} (\bibinfo
	{year} {2019})}\BibitemShut {NoStop}%
\bibitem [{\citenamefont {Matityahu}\ \emph {et~al.}(2016)\citenamefont
	{Matityahu}, \citenamefont {Utsumi}, \citenamefont {Aharony}, \citenamefont
	{Entin-Wohlman},\ and\ \citenamefont {Balseiro}}]{matityahu2016spin}%
\BibitemOpen
\bibfield  {author} {\bibinfo {author} {\bibfnamefont {Shlomi}\ \bibnamefont
		{Matityahu}}, \bibinfo {author} {\bibfnamefont {Yasuhiro}\ \bibnamefont
		{Utsumi}}, \bibinfo {author} {\bibfnamefont {Amnon}\ \bibnamefont {Aharony}},
	\bibinfo {author} {\bibfnamefont {Ora}\ \bibnamefont {Entin-Wohlman}}, \ and\
	\bibinfo {author} {\bibfnamefont {Carlos~A}\ \bibnamefont {Balseiro}},\
}\bibfield  {title} {\enquote {\bibinfo {title} {Spin-dependent transport
		through a chiral molecule in the presence of spin-orbit interaction and
		nonunitary effects},}\ }\href@noop {} {\bibfield  {journal} {\bibinfo
	{journal} {Physical Review B}\ }\textbf {\bibinfo {volume} {93}},\ \bibinfo
{pages} {075407} (\bibinfo {year} {2016})}\BibitemShut {NoStop}%
\bibitem [{\citenamefont {Medina}\ \emph {et~al.}(2015)\citenamefont {Medina},
	\citenamefont {Gonz\'alez-Arraga}, \citenamefont {Finkelstein-Shapiro},
	\citenamefont {Berche},\ and\ \citenamefont {Mujica}}]{medina2015continuum}%
\BibitemOpen
\bibfield  {author} {\bibinfo {author} {\bibfnamefont {Ernesto}\ \bibnamefont
		{Medina}}, \bibinfo {author} {\bibfnamefont {Luis~A}\ \bibnamefont
		{Gonz\'alez-Arraga}}, \bibinfo {author} {\bibfnamefont {Daniel}\ \bibnamefont
		{Finkelstein-Shapiro}}, \bibinfo {author} {\bibfnamefont {Bertrand}\
		\bibnamefont {Berche}}, \ and\ \bibinfo {author} {\bibfnamefont {Vladimiro}\
		\bibnamefont {Mujica}},\ }\bibfield  {title} {\enquote {\bibinfo {title}
		{Continuum model for chiral induced spin selectivity in helical molecules},}\
}\href@noop {} {\bibfield  {journal} {\bibinfo  {journal} {The Journal of
		Chemical Physics}\ }\textbf {\bibinfo {volume} {142}},\ \bibinfo {pages}
{194308} (\bibinfo {year} {2015})}\BibitemShut {NoStop}%
\bibitem [{\citenamefont {N{\"u}renberg}\ and\ \citenamefont
	{Zacharias}(2019)}]{nurenberg2019evaluation}%
\BibitemOpen
\bibfield  {author} {\bibinfo {author} {\bibfnamefont {Daniel}\ \bibnamefont
		{N{\"u}renberg}}\ and\ \bibinfo {author} {\bibfnamefont {Helmut}\
		\bibnamefont {Zacharias}},\ }\bibfield  {title} {\enquote {\bibinfo {title}
		{Evaluation of spin-flip scattering in chirality-induced spin selectivity
			using the riccati equation},}\ }\href@noop {} {\bibfield  {journal} {\bibinfo
		{journal} {Physical Chemistry Chemical Physics}\ }\textbf {\bibinfo {volume}
		{21}},\ \bibinfo {pages} {3761--3770} (\bibinfo {year} {2019})}\BibitemShut
{NoStop}%
\bibitem [{\citenamefont {Dalum}\ and\ \citenamefont
	{Hedeg{\aa}rd}(2019)}]{dalum2019theory}%
\BibitemOpen
\bibfield  {author} {\bibinfo {author} {\bibfnamefont {Sakse}\ \bibnamefont
		{Dalum}}\ and\ \bibinfo {author} {\bibfnamefont {Per}\ \bibnamefont
		{Hedeg{\aa}rd}},\ }\bibfield  {title} {\enquote {\bibinfo {title} {Theory of
			chiral induced spin selectivity},}\ }\href@noop {} {\bibfield  {journal}
	{\bibinfo  {journal} {Nano Letters}\ }\textbf {\bibinfo {volume} {19}},\
	\bibinfo {pages} {5253--5259} (\bibinfo {year} {2019})}\BibitemShut {NoStop}%
\bibitem [{\citenamefont {B{\"u}ttiker}(1986)}]{buttiker1986role}%
\BibitemOpen
\bibfield  {author} {\bibinfo {author} {\bibfnamefont {M}~\bibnamefont
		{B{\"u}ttiker}},\ }\bibfield  {title} {\enquote {\bibinfo {title} {Role of
			quantum coherence in series resistors},}\ }\href@noop {} {\bibfield
	{journal} {\bibinfo  {journal} {Physical Review B}\ }\textbf {\bibinfo
		{volume} {33}},\ \bibinfo {pages} {3020} (\bibinfo {year}
	{1986})}\BibitemShut {NoStop}%
\bibitem [{\citenamefont {S{\'a}nchez}\ and\ \citenamefont
	{B{\"u}ttiker}(2004)}]{sanchez2004magnetic}%
\BibitemOpen
\bibfield  {author} {\bibinfo {author} {\bibfnamefont {David}\ \bibnamefont
		{S{\'a}nchez}}\ and\ \bibinfo {author} {\bibfnamefont {Markus}\ \bibnamefont
		{B{\"u}ttiker}},\ }\bibfield  {title} {\enquote {\bibinfo {title}
		{Magnetic-field asymmetry of nonlinear mesoscopic transport},}\ }\href@noop
{} {\bibfield  {journal} {\bibinfo  {journal} {Physical Review Letters}\
	}\textbf {\bibinfo {volume} {93}},\ \bibinfo {pages} {106802} (\bibinfo
	{year} {2004})}\BibitemShut {NoStop}%
\bibitem [{\citenamefont {Datta}(1997)}]{datta1997electronic}%
\BibitemOpen
\bibfield  {author} {\bibinfo {author} {\bibfnamefont {Supriyo}\ \bibnamefont
		{Datta}},\ }\href@noop {} {\emph {\bibinfo {title} {Electronic transport in
			mesoscopic systems}}}\ (\bibinfo  {publisher} {Cambridge University Press},\
\bibinfo {year} {1997})\BibitemShut {NoStop}%
\bibitem [{\citenamefont {Ben-Chorin}\ \emph {et~al.}(1994)\citenamefont
	{Ben-Chorin}, \citenamefont {M{\"o}ller},\ and\ \citenamefont
	{Koch}}]{ben1994nonlinear}%
\BibitemOpen
\bibfield  {author} {\bibinfo {author} {\bibfnamefont {M}~\bibnamefont
		{Ben-Chorin}}, \bibinfo {author} {\bibfnamefont {F}~\bibnamefont
		{M{\"o}ller}}, \ and\ \bibinfo {author} {\bibfnamefont {F}~\bibnamefont
		{Koch}},\ }\bibfield  {title} {\enquote {\bibinfo {title} {Nonlinear
			electrical transport in porous silicon},}\ }\href@noop {} {\bibfield
	{journal} {\bibinfo  {journal} {Physical Review B}\ }\textbf {\bibinfo
		{volume} {49}},\ \bibinfo {pages} {2981} (\bibinfo {year}
	{1994})}\BibitemShut {NoStop}%
\bibitem [{\citenamefont {Middleton}\ and\ \citenamefont
	{Wingreen}(1993)}]{middleton1993collective}%
\BibitemOpen
\bibfield  {author} {\bibinfo {author} {\bibfnamefont {A~Alan}\ \bibnamefont
		{Middleton}}\ and\ \bibinfo {author} {\bibfnamefont {Ned~S}\ \bibnamefont
		{Wingreen}},\ }\bibfield  {title} {\enquote {\bibinfo {title} {Collective
			transport in arrays of small metallic dots},}\ }\href@noop {} {\bibfield
	{journal} {\bibinfo  {journal} {Physical Review Letters}\ }\textbf {\bibinfo
		{volume} {71}},\ \bibinfo {pages} {3198} (\bibinfo {year}
	{1993})}\BibitemShut {NoStop}%
\bibitem [{\citenamefont {Michaeli}\ \emph {et~al.}(2019)\citenamefont
	{Michaeli}, \citenamefont {Beratan}, \citenamefont {Waldeck},\ and\
	\citenamefont {Naaman}}]{michaeli2019voltage}%
\BibitemOpen
\bibfield  {author} {\bibinfo {author} {\bibfnamefont {Karen}\ \bibnamefont
		{Michaeli}}, \bibinfo {author} {\bibfnamefont {David~N}\ \bibnamefont
		{Beratan}}, \bibinfo {author} {\bibfnamefont {David~H}\ \bibnamefont
		{Waldeck}}, \ and\ \bibinfo {author} {\bibfnamefont {Ron}\ \bibnamefont
		{Naaman}},\ }\bibfield  {title} {\enquote {\bibinfo {title} {Voltage-induced
			long-range coherent electron transfer through organic molecules},}\
}\href@noop {} {\bibfield  {journal} {\bibinfo  {journal} {Proceedings of the
		National Academy of Sciences}\ }\textbf {\bibinfo {volume} {116}},\ \bibinfo
{pages} {5931--5936} (\bibinfo {year} {2019})}\BibitemShut {NoStop}%
\bibitem [{\citenamefont {Kiran}\ \emph {et~al.}(2016)\citenamefont {Kiran},
	\citenamefont {Mathew}, \citenamefont {Cohen}, \citenamefont
	{Hern{\'a}ndez~Delgado}, \citenamefont {Lacour},\ and\ \citenamefont
	{Naaman}}]{kiran2016helicenes}%
\BibitemOpen
\bibfield  {author} {\bibinfo {author} {\bibfnamefont {Vankayala}\
		\bibnamefont {Kiran}}, \bibinfo {author} {\bibfnamefont {Shinto~P}\
		\bibnamefont {Mathew}}, \bibinfo {author} {\bibfnamefont {Sidney~R}\
		\bibnamefont {Cohen}}, \bibinfo {author} {\bibfnamefont {Irene}\ \bibnamefont
		{Hern{\'a}ndez~Delgado}}, \bibinfo {author} {\bibfnamefont {J{\'e}r{\^o}me}\
		\bibnamefont {Lacour}}, \ and\ \bibinfo {author} {\bibfnamefont {Ron}\
		\bibnamefont {Naaman}},\ }\bibfield  {title} {\enquote {\bibinfo {title}
		{Helicenes--a new class of organic spin filter},}\ }\href@noop {} {\bibfield
	{journal} {\bibinfo  {journal} {Advanced Materials}\ }\textbf {\bibinfo
		{volume} {28}},\ \bibinfo {pages} {1957--1962} (\bibinfo {year}
	{2016})}\BibitemShut {NoStop}%
	\bibitem [{\citenamefont {Yang}\ \emph
	{et~al.}(2019{\natexlab{b}})\citenamefont {Yang}, \citenamefont {van~der
		Wal},\ and\ \citenamefont {van Wees}}]{yang2019detecting}%
\BibitemOpen
\bibfield  {author} {\bibinfo {author} {\bibfnamefont {Xu}~\bibnamefont
		{Yang}}, \bibinfo {author} {\bibfnamefont {Caspar~H}\ \bibnamefont {van~der
			Wal}}, \ and\ \bibinfo {author} {\bibfnamefont {Bart~J}\ \bibnamefont {van
			Wees}},\ }\bibfield  {title} {\enquote {\bibinfo {title} {Detecting chirality
			in two-terminal electronic devices},}\ }\href@noop {} {\bibfield  {journal}
	{\bibinfo  {journal} {arXiv preprint}\ ,\ \bibinfo {pages} {1912.09085}}
	(\bibinfo {year} {2019}{\natexlab{b}})}\BibitemShut {NoStop}%
\end{thebibliography}
\end{document}